\documentclass[sigconf,nonacm]{acmart}
\usepackage{multirow}
\usepackage{float}
\begin{document}
\title{MM-SVGEdit: A Multimodal-Driven SVG Editing for UI Design}

\author{Shibo Yang}
\authornote{Both authors contributed equally to this research.}
\email{bryan.yang.shibo@centralepekin.cn}
\affiliation{%
  \institution{Beihang University}
  \city{Beijing}
  \country{China}}

\author{Yuqing Gao}
\authornotemark[1]
\email{gyq0808g@163.com}
\affiliation{%
  \institution{Beihang University}
  \city{Beijing}
  \country{China}}

\author{Zipeng Liu}
\email{zipeng@buaa.edu.cn}
\affiliation{%
  \institution{Beihang University}
  \city{Beijing}
  \country{China}}

\renewcommand{\shortauthors}{Yang et al.}

\begin{teaserfigure}
  \centering
  \includegraphics[width=\textwidth]{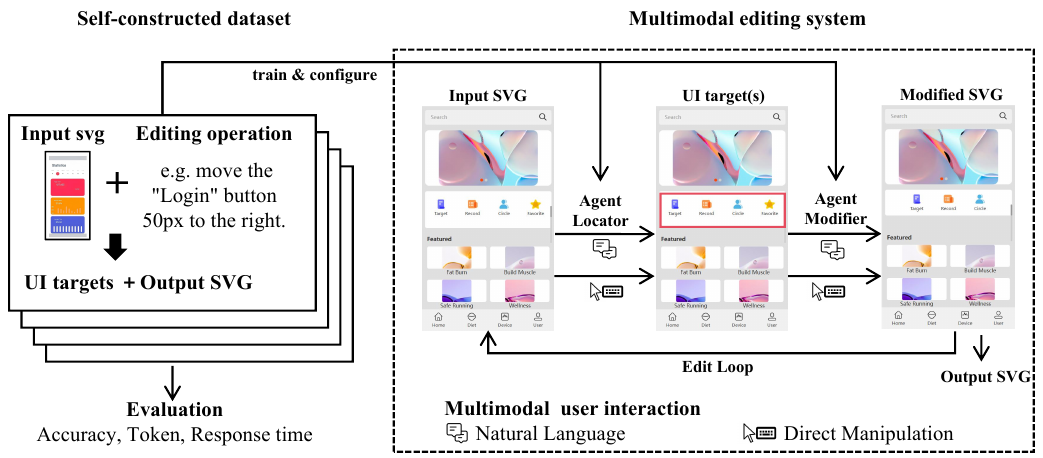}
  \caption{Overview of the multimodal editing system. 
For the multimodal editing system, starting from an SVG preview, we propose a two-stage 
workflow comprising visual grounding and interactive modification. Both stages support 
dual interaction modalities—direct manipulation via mouse/keyboard and LLM-based agent 
authoring via natural language, which enables accurate visual grounding and modification of target elements within the SVG preview. The example demonstrates the deletion of the \textit{favorite} button and uniform spacing among the remaining elements.
To effectively train and evaluate this multimodal editing system, we construct a 
comprehensive SVG-based UI element dataset by converting existing bitmap-based UI datasets, and further augment it with custom editing operation annotations to create a visual grounding and modification dataset.}
  \Description{Complete overview showing multimodal input through Agent Locator and output through Agent Modifier.}
  \label{fig:svg-editing-scheme}
\end{teaserfigure}

\begin{abstract}

In the field of UI design, Scalable Vector Graphics (SVG) is widely used as a design medium. However, traditional SVG editing techniques have high entry barriers and require cumbersome manual iteration, while LLM-based editing solutions suffer from low accuracy and poor user controllability. To address these issues, we propose MM-SVGEdit, a multimodal-driven SVG editing approach that integrates traditional SVG editing and LLM-based methods. We introduce a two-stage strategy — visual grounding first, then modification — in which both stages support two interaction modalities: natural language instructions and direct manipulation (mouse and keyboard). 
We trained and evaluated MM-SVGEdit on a self-constructed dataset of 14,476 question-answer pairs generated from UIs, covering 11 types of editing operations on both single and multiple UI targets. The results show that MM-SVGEdit improves SVG editing accuracy, efficiency, and user-perceived control while reducing token consumption and response time.

\end{abstract}

\ccsdesc[500]{Human-centered computing~Human computer interaction (HCI)}
\ccsdesc[500]{Human-centered computing~Interaction paradigms}
\ccsdesc[500]{Human-centered computing~User interface management systems}

\keywords{Human-AI Interaction, UI Design, Large Language Model, Multimodal Interaction}
\maketitle

\section{Introduction}
Traditional SVG editing tools are built around manual workflows that assume significant professional expertise.
Desktop tools such as Adobe Illustrator and Inkscape demand extensive design expertise and familiarity with complex toolsets, while web-based libraries such as Snap.svg and D3.js additionally require proficiency in programming and SVG syntax.
This high barrier excludes ordinary users and business personnel from efficient SVG creation. Even for professional users, tedious operations such as node adjustment, path editing, and parameter configuration severely restrict agility and creativity.

This work focuses on structured, browser-renderable SVG artifacts for UIs. Rather than replacing full-featured design environments such as Figma or Sketch, it supports localized, code-backed editing for lightweight prototyping and web-oriented workflows.

The rapid development of large language models (LLMs) has provided a new way to bridge this gap~\cite{he2024llms}. 
LLM-driven SVG editing has evolved rapidly along two directions: text-to-graphics generation, where models produce SVG code directly from natural language descriptions, and instruction-based modification, where frameworks support intelligent editing via text commands.
However, existing methods still suffer from critical limitations~\cite{malashenko2025leveraging}. First, relying solely on text leads to ambiguous spatial specifications and imprecise element targeting — for instance, the instruction ``move the icon slightly to the right of the button'' leaves critical details unspecified, such as which icon, which button, and the exact offset distance.
Second, the black-box nature of LLM generation may cause syntax errors, structural distortion, and insufficient detail control, depriving users of fine-grained direct control.

These limitations point to a fundamental gap: traditional manual interaction and LLM-based editing have largely remained isolated, and neither alone can deliver both the precision and the intelligence that SVG editing demands. Bridging this gap — combining the accuracy of direct manipulation with the flexibility of LLM-driven editing — represents the key to overcoming these bottlenecks. Nevertheless, such integration remains underexplored, with open challenges in aligning text commands with spatial intent, ensuring controllability of LLM outputs, and designing transparent collaborative workflows. Therefore, a multimodal, LLM-enhanced SVG editing system is both timely and necessary~\cite{ouyang2022training}.

We propose an SVG editing technique that integrates traditional mouse-keyboard editing and LLM-based editing, enabling users to select the suitable interaction modality for each task.
Central to this approach is a two-stage strategy: a visual grounding stage that locates the target element, followed by a modification stage that applies the intended edits, with both stages supporting multimodal interaction.
To support editing instructions with natural language, we developed two LLM agents, \textbf{Agent Locator} for the visual grounding stage, and \textbf{Agent Modifier} for the modification stage, as shown in Figure~\ref{fig:svg-editing-scheme}.  Both are trained and configured on a self-constructed dataset consisting of UI images in the SVG format, representative editing operations, UI element targets, and the modified SVG as ground-truth results.  Specifically, Agent Locator is finetuned based on an open-source lightweight LLM, while Agent Modifier is configured with a function calling mechanism.
Experiments on this dataset demonstrate that MM-SVGEdit improves SVG editing accuracy and efficiency, while also enhancing user-perceived control over the editing process — all while maintaining token consumption and response time within acceptable bounds, confirming that the integration of LLM components introduces minimal computational overhead.


In summary, we make the following contributions.
\begin{enumerate}
    \item We propose a two-stage strategy — visual grounding first, then modification — in which both stages support multimodal interaction via mouse-keyboard manipulation and natural language instructions.
    \item We develop two LLM agents, Agent Locator and Agent Modifier, along with a self-constructed dataset of UI SVGs and representative editing operations, enabling systematic training and evaluation of both agents.
    \item We implement MM-SVGEdit, a multimodal SVG editing system for UI design that integrates the proposed strategy and agents, demonstrating improvements in editing accuracy, efficiency, and user-perceived control with minimal computational overhead.
\end{enumerate}

\section{Related Work}

\subsection{Traditional SVG Editing Technologies}

Traditional SVG editing commonly relies on manual operations in professional tools.
Before the rise of AI-assisted design, they represented the primary approach to SVG creation.
These technologies center on desktop tools and development libraries. Their use requires design expertise or coding skills, resulting in high entry barriers and low operational efficiency.

In terms of editing tools, traditional SVG editing mainly relies on professional desktop software such as Adobe Illustrator and Inkscape. These software programs are equipped with comprehensive and fine-grained SVG editing functions and serve as core tools for professional designers. In addition, development libraries including Snap.svg and SVG.js are also widely used in traditional SVG editing, enabling developers to manually edit SVG code to meet customized needs. 
However, both categories depend entirely on manual effort — whether path editing and node adjustment in desktop tools, or code debugging in development libraries, which not only leads to cumbersome operations for experts but also excludes ordinary users and business personnel from efficient SVG creation~\cite{robinson2019sketch2code}.

In terms of application scenarios, traditional SVG editing technologies are widely used in professional design and front-end development fields, but have obvious limitations: even for professional users, tedious manual operations severely restrict the flexibility and innovation of creation, leading to the dilemma of ``tool-constrained thinking''~\cite{feng2021auto}. In addition, traditional SVG editing lacks intelligent assistance capabilities and cannot support the automatic generation or modification of SVG content based on natural language instructions, making it difficult to meet the demand for efficient SVG creation in modern design scenarios.

\subsection{AI-Driven SVG Editing Technologies}

AI-assisted visual editing has developed across both raster-image and vector-graphics representations. In raster-image research, XMC-GAN improves text-to-image generation through cross-modal contrastive learning~\cite{zhang2021cross}; Tov et al. study StyleGAN inversion for real-image manipulation~\cite{tov2021designing}; and Stap et al. introduce text-conditioned image generation and semantic facial manipulation~\cite{stap2020conditional}. Diffusion-based methods subsequently improved image-synthesis fidelity and controllability~\cite{saharia2022photorealistic,huang2025diffusion}, while SmartEdit and InstructPix2Pix explored instruction-based raster-image editing~\cite{huang2024smartedit,brooks2023instructpix2pix}. Other raster methods support localized semantic painting, zero-shot image translation, and spatial conditioning~\cite{andonian2021paint,parmar2023zero,zhang2023adding}. These methods do not directly manipulate SVG code, but provide relevant paradigms for language-guided visual editing.

Recent SVG-specific work instead exploits the structured textual representation of vector graphics. Cai et al. investigate visual understanding through SVG representations~\cite{cai2025investigation}, while VDLM introduces an intermediate textual representation for vector-graphics reasoning~\cite{wang2024visually}.
Chat2SVG first uses an LLM to generate a semantically structured SVG template. Image diffusion models then produce a more detailed raster target, after which a separate dual-stage SVG optimization process refines path representations in latent space and adjusts control-point coordinates to match the target~\cite{wu2025chat2svg}.
Visual ChatGPT integrates ChatGPT with multiple visual foundation models, enabling users to submit images and language instructions, invoke multi-step visual reasoning and editing, and provide feedback for correction~\cite{wu2023visual}. It is a general visual-model orchestration framework rather than an SVG-specific editing method.

Xu et al. translate language instructions into geometric constraints to create spatial variations of segmented icon images~\cite{xu2024creating}. Xu and Wall evaluate zero-shot LLMs on ten low-level visual analytic tasks over SVG visualizations and find that performance varies substantially across tasks~\cite{xu2024exploring}. Together, these studies show that language interfaces can support spatial icon editing and selected SVG analytic tasks. The former focuses on spatial relations among segmented icon components, whereas the latter reports uneven performance across tasks, particularly for operations requiring numerical computation.

Another important research direction focuses on the ``design-to-code'' paradigm, which enables automatic front-end code generation from UI screenshots, design drafts, or other visual representations. This research has 
attracted considerable attention from the community due to its potential to lower the 
barrier for software development. Representative works include pix2code~\cite{beltramelli2018pix2code}, Alibaba's imgcook~\cite{Imgcook}, and Design2Code, a benchmark of 484 real-world webpages for evaluating screenshot-to-code generation by multimodal LLMs~\cite{si2025design2code}. These works demonstrate the feasibility of automating front-end implementation from visual designs. Design2Code further reports that current multimodal models still struggle to recover visual elements and generate accurate layouts. Overall, this line of research focuses primarily on code generation from visual designs rather than iterative editing of vector-graphic content.

\subsection{Datasets for SVG Editing}
The training and evaluation of AI-driven SVG editing models depend on dedicated datasets that provide vector graphics paired with editing instructions.
Existing UI design datasets have achieved significant scale. The Rico dataset~\cite{deka2017Rico} 
is a widely-used resource in UI screen understanding, containing over 66k unique mobile UI 
screens with their view hierarchies. Subsequent work has augmented it with 500k human 
annotations for icon shape and semantic recognition, and established associations between 
common UI elements and their text labels. Similarly, mrtoy~\cite{mrtoy2023mobileui} is another large-scale mobile 
UI dataset designed for element detection and recognition. However, these traditional UI 
datasets are primarily bitmap-based, which poses fundamental limitations for SVG editing 
tasks due to the difficulty of precise element manipulation and property modification 
in raster formats.

In recent years, dedicated SVG editing datasets have emerged to address the need for 
vector graphics editing: the SAgoge dataset~\cite{wang2025internsvg} constructs 16 million 
multimodal SVG samples enabling unified training for SVG understanding, editing, and 
generation; the VectorEdits dataset~\cite{kuchavr2025vectoredits} provides more than 270,000 
SVG-editing instruction pairs, addressing the limitation of single task type in early 
datasets; SVGEditBench builds a lightweight evaluation benchmark based on Twemoji data 
for standardized assessments of six major LLM-driven SVG editing tasks~\cite{nishina2025svgeditbench}; 
the SVGenius dataset~\cite{chen2025svgenius} stratifies real-world editing queries by 
complexity to enable systematic performance evaluation. Nevertheless, despite their 
contributions, these existing SVG datasets face a critical limitation: they are primarily 
designed for general-purpose vector graphics editing rather than specialized UI-specific 
editing tasks. This results in a significant gap between the dataset content and real-world 
UI design requirements, making it difficult to apply these datasets directly to professional 
UI editing scenarios.

To address these limitations, we construct a dedicated SVG-based UI editing dataset by 
systematically converting existing bitmap-based UI datasets (Rico~\cite{deka2017Rico}, mrtoy~\cite{mrtoy2023mobileui}) into vector 
graphics format. We further augment this dataset with fine-grained element annotations 
and high-quality instruction-action pairs tailored specifically for UI editing tasks, 
enabling comprehensive training and evaluation of UI-focused SVG editing models.

\section{Multimodal Editing System}

\subsection{Multimodal Scheme}
\label{subsec:multimodal}
This section presents the core methods of the proposed system, which integrates LLMs with traditional editing techniques to enable precise multimodal interaction on SVG images, supporting iterative UI design refinement, as illustrated on the right side of Figure~\ref{fig:svg-editing-scheme}. The methodology decomposes SVG refinement into two key stages: visual grounding and multimodal interactive modification. Both stages support multiple interaction modalities, allowing users to select the one best suited to the task at hand.

The core innovation lies in a framework that tightly integrates traditional mouse/keyboard editing with LLM-driven natural language editing within a unified environment. Unlike existing tools that treat manual and AI-driven editing as independent workflows, the proposed system allows users to seamlessly switch between or combine modalities. For instance, a user can manually reposition a button via mouse drag, then issue a natural language instruction (e.g., ``set the color of all buttons to blue'') to batch-modify multiple elements all within a single session. This integration fills a critical gap in current UI design tools, which largely lack support for multimodal editing and fine-grained iterative SVG refinement.

These modes are not intended as four independent features to be invoked in every editing task. They provide complementary interaction strategies for different trade-offs: natural-language input is suited to semantic or multi-element changes, whereas direct manipulation supports precise local refinements. Hybrid modes combine these strengths when target selection or the intended modification requires human correction. Specifically, there are four complementary interaction modes in an editing round.

\begin{enumerate}
    \item \textbf{Manual interaction only:} Direct mouse/keyboard operations for fine-grained positional and style adjustments. \textit{Example:} A designer drags a button element to reposition it on the canvas with sub-pixel precision, with the editing engine adopting an incremental update mechanism to ensure real-time responsiveness.
    
    \item \textbf{LLM-only editing:} The system performs semantic modifications from unconstrained natural language instructions. \textit{Example:} A user inputs ``change all button backgrounds to blue and add a 2px border.'' The LLM converts the request into structured function calls, which the SVG editing engine applies uniformly to all buttons.
    
    \item \textbf{Manual selection + LLM modification:} Users employ box selection via mouse to precisely identify multiple target elements, then issue natural language commands for batch processing. \textit{Example:} A designer box-selects five text elements scattered across the design, then instructs ``unify font size to 16px and set line-height to 1.5,'' which the LLM executes only on the selected elements.
    
    \item \textbf{LLM pre-selection + manual refinement:} The fine-tuned model automatically identifies and highlights candidate elements (e.g., all card components); users then selectively verify, remove, or adjust candidates via mouse clicks, exercising explicit human oversight over model outputs. \textit{Example:} The model marks all detected navigation cards; the designer clicks to deselect false positives, then applies a unified styling rule only to confirmed cards.
\end{enumerate}

The system integrates these modes into a single SVG editing workflow. It first obtains candidate target elements through direct selection or Agent Locator; users may inspect and revise the selection before Agent Modifier translates the editing instruction into constrained function calls. The editing engine applies these calls only to the confirmed target elements, preserving unaffected SVG content. This workflow connects element localization, human correction, and localized structured modification in an iterative editing round.

\subsection{Agent Locator}

The Agent Locator serves as a critical component for understanding UI-oriented editing instructions and precisely identifying target elements within SVG graphics. It takes as input an SVG file and a natural language editing instruction, and outputs the ID(s) of the target UI element(s) to be manipulated. To enhance the ability of large language models to understand UI-oriented instructions and accurately locate target elements in SVG, we construct a dedicated instruction tuning dataset that provides high-quality and task-specific data support for model fine-tuning. The dataset is structured in the form of question-answer (QA) pairs, where each sample consists of three core fields: question (natural language instruction), answer-id (target element ID), and source (the corresponding SVG file). This structured format of instruction-description, target-ID, and data-source enables the model to effectively associate user instructions, target elements, and the original SVG context during training, improving the accuracy and relevance of learning. 

The dataset contains 14,476 high-quality QA pairs covering 11 typical UI editing operations, including element alignment, synchronous movement, uniform sizing, batch styling, layout arrangement, layer management, component grouping, batch text modification, batch visibility control, intelligent layout, and advanced conditional operations. These operations cover common editing requirements in mobile UI design, ranging from basic manipulations to complex semantic interactions, and are used to train the model to fully grasp the semantic logic and execution scope of various UI instructions and establish a precise mapping between natural language commands and SVG target elements. The detailed information of the dataset will be elaborated in Section~\ref{sec:dataset}. 

To improve the accuracy of UI element visual grounding, we perform parameter-efficient fine-tuning based on an open-source lightweight LLM, Qwen2.5-7B-Instruct model~\cite{qwen2.5}. It has strong instruction-following capability and structured data understanding, making it well-suited for domain-specific tasks such as visual grounding of SVG elements. Its advantages in long-text processing, multilingual understanding, and instruction execution provide a solid foundation for UI semantic parsing and target element positioning. 

For fine-tuning, we adopt LoRA (Low-Rank Adaptation)~\cite{hu2022lora}, a parameter-efficient technique that avoids updating the full set of model parameters. By inserting trainable low-rank matrices into the attention layers of the original model, LoRA achieves performance close to full-parameter fine-tuning while training only a small number of parameters. This greatly reduces GPU memory consumption and training costs, preserves the original capabilities of the pre-trained model, and mitigates overfitting, ensuring efficient and stable fine-tuning.
The use of a lightweight model and LoRA technique allows us to perform finetuning efficiently on a machine with a single NVIDIA RTX 4090 (24GB) GPU.

In the training procedure, we select 5,000 high-quality samples from the 14,476 QA pairs and split them at a ratio of 9:1 into training and test sets. Due to the adoption of efficient LoRA fine-tuning, the model achieves satisfactory performance with 5,000 samples, and further training shows no significant decrease in loss. Therefore, we determine 5,000 samples as the final training set scale. In each iteration, the model receives structured SVG information, a natural language UI editing instruction, and the correct target element ID as supervision. The model outputs the element ID it identifies as the target.

During training, the loss is computed by comparing the predicted ID with the ground-truth ID, and the parameters of the LoRA branches are updated via backpropagation to gradually reduce the Cross-Entropy Loss. After multiple iterations until the loss stabilizes and converges, we obtain the final fine-tuned Qwen2.5-7B-Instruct model, which is then evaluated on the held-out test set to verify its generalization ability on unseen UI editing instructions.

Experimental results demonstrate that the proposed dataset and fine-tuning strategy significantly improve the model's UI semantic understanding and element visual grounding accuracy. The fine-tuned model substantially outperforms the original pre-trained version, effectively resolves ambiguities in natural language instructions, and accurately identifies target elements under various complex scenarios. It also achieves performance comparable to commercial large language models such as GPT-4o and DeepSeek, especially in single-target UI element visual grounding. This provides a stable and precise foundation for the subsequent function-call-based SVG editing module in the overall system.

\subsection{Agent Modifier}

The Agent Modifier is responsible for executing precise UI editing operations based on identified target elements. It takes as input an SVG file, a natural language editing instruction, and the target element ID(s), and outputs the modified SVG file with the requested edits applied. To address the issues of uncertainty and inaccuracy when large language models (LLMs) manipulate SVG elements, this module adopts a function calling mechanism that decouples intent parsing from operation execution. The module comprises two key components: the Specification Module, which constrains model outputs through standardized function definitions, and the Execution Module, which performs localized modifications of UI elements, enabling stable and controllable UI editing while ensuring the precision and efficiency of operations.

Based on the visual grounding from the first stage, which can be achieved either through direct manipulation by selecting elements or using the Agent Locator model trained for UI element visual grounding, the target element ID is first obtained to enhance the accuracy of subsequent modification operations. This enables structured control over SVG files via a function calling mechanism. The Specification Module defines standardized function specifications including function names, parameter lists, and functional descriptions to assist the LLM in understanding UI editing instructions and generating compliant calling parameters. Core function categories are defined in the mechanism to handle common UI editing operations: element extraction for obtaining panoramic element information, attribute modification for editing operations such as color adjustment and position movement, element deletion for removing redundant UI components, and element addition for inserting new UI components. The specification design adheres to the principles of ``least privilege'' and ``single responsibility'': for text modification, only element ID and text content are allowed to be passed in; when adding elements, legal tags and attributes must be provided; all parameters must comply with the parameter specifications defined in the Specification Module to avoid parsing failures caused by invalid parameters.

To improve the accuracy of target element editing while enabling direct user participation in LLM-driven editing, the module adopts a two-step interaction workflow of ``function calling generation — operation execution''.

\textbf{Step 1: Function Calling Generation.} After the user inputs a natural language instruction, the target element ID is obtained from the visual grounding stage, ensuring that function calling code is generated only for the identified target element without modifying any other elements, thereby reducing the possibility of erroneous modifications. For SVG files without complete element metadata, the Specification Module guides the model to first extract metadata of all editable elements. After confirming the target elements, the SVG context, editing instructions, and specification constraints are input into the LLM. Under the guidance of contextualized prompts and standardized specifications, compliant function calling code is generated. For example, to move the red button in the top-right corner, the corresponding function call with movement parameters is generated according to specification.

\textbf{Step 2: Structured Operation Execution.} After obtaining the function calling code, the elements that will be modified are first presented to the user for confirmation. If users are unsatisfied, they can refine the input through further modifications to adjust the LLM output. Upon confirmation, the Execution Module parses the function calling information, executes the corresponding SVG modification functions, and verifies the format compliance of the modified SVG. Finally, the modified SVG and operation logs are returned to the front-end to complete an editing process.

\begin{figure}[h]
  \centering
  \includegraphics[width=\linewidth]{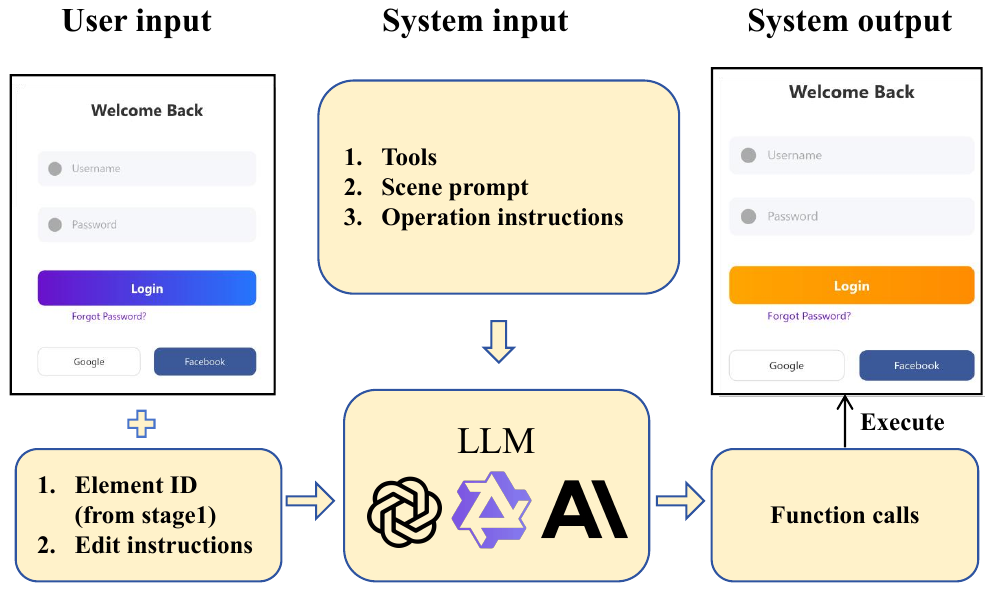}
  \caption{Design scheme of the function calling mechanism}
  \Description{Design Scheme of Agent Modifier}
\end{figure}

The core advantage of the two-step workflow lies in effectively reducing erroneous modifications by LLMs, standardizing structured outputs, and enhancing user control throughout the editing process. First, by relying on accurate element ID localization, positioning errors caused by text-only reasoning are substantially mitigated. Compared to direct SVG modification, where LLMs frequently encounter issues such as missing elements and chaotic modifications, the function calling mechanism leverages the mandatory standardization of the Specification Module and contextualized prompts to reduce task complexity: the model only needs to identify relevant parameters rather than directly modify the entire SVG. The standardized output ensures consistent SVG editing across multiple modifications, preventing LLM's overwriting regeneration and effectively preserving the original SVG style.

Most importantly, unlike traditional LLM-based editing where user control is limited to prompt refinement, this method allows users to directly influence model outputs through stricter specification function definitions. During the editing process, users can monitor LLM behavior through UI feedback or by directly inspecting the function calling code, enabling further interaction with the model and active participation in the editing workflow. This comprehensively improves the precision and reliability of SVG editing, significantly reducing LLM output instability.

The function calling approach significantly enhances the efficiency of SVG editing in terms of both computational cost and user experience. By constraining model outputs through the Specification Module and delegating execution to the lightweight Execution Module, token consumption is substantially reduced compared to end-to-end generation approaches. This reduction in token usage directly translates to faster inference speeds and lower latency, enabling rapid real-time interaction with users. Furthermore, the structured nature of function calling eliminates unnecessary model computation, allowing smaller and lower-precision models to achieve comparable performance on SVG editing tasks. The overall editing efficiency is substantially improved, as users experience faster response times and can iterate through multiple editing cycles more rapidly, dramatically enhancing the interactivity and usability of the UI editing system.

\subsection{System Architecture and Interface}
The core design idea of this system is to integrate traditional editing tools with AI editing tools to construct a multimodal interactive editing system. The system adopts a four-layer architecture: User Interaction Layer, Intelligent Reasoning Layer, Editing Engine Layer, and Data Persistence Layer, decomposing UI element recognition and UI element modification into two steps. It integrates a bimodal interaction mode of natural language input and mouse drag-and-drop operations, supporting multiple interaction scenarios including independent mouse/keyboard operation, independent LLM processing, and collaborative work of the two modalities. 

\subsubsection{Layers Design}

The subsequent sections detail the specific implementation of multimodal SVG editing technology through a four-layer architecture, where each layer integrates with the overall system pipeline illustrated in Figure 1 to enable seamless user-system collaboration.

\textbf{User Interaction Layer:} This layer bridges users and the system through a natural language input box, direct manipulation on the SVG canvas, and a real-time preview panel. It supports four modes: mouse-and-keyboard-only editing, LLM-only editing, direct-manipulation-assisted LLM editing, and LLM-assisted direct manipulation. The layer captures both direct manipulation and natural language instructions as expressions of user intent. For hybrid tasks, a user can select elements precisely with the mouse and then issue a natural language command for batch modification.

Crucially, this layer provides bidirectional feedback between user intent and system output. It receives results from the Intelligent Reasoning Layer and the Editing Engine Layer, then relays them to users in real time. For example, after a user requests ``move the three left buttons to the right,'' the interface highlights the identified elements for confirmation before modification. If the intent is ambiguous, the system requests clarification through additional text or direct mouse selection.

\textbf{Intelligent Reasoning Layer:} The Intelligent Reasoning Layer deploys a fine-tuned Qwen2.5-7B-Instruct model, which is used to accurately locate target elements and then parse natural language into SVG operation parameters through the Function Calling mechanism. This layer designs a UI semantics-function mapping module, establishing a three-level mapping relationship between user instruction keywords, SVG element attributes, and modification functions, eliminating the ambiguity of natural language and improving the instruction parsing accuracy to 92\% on a mixed evaluation set containing both single-target and multi-target instructions. The Intelligent Reasoning Layer can accurately understand the semantics of user natural language instructions, clarify editing goals and operation types, accurately identify target elements in SVG documents according to the parsing results, and generate specific SVG operation instructions such as moving, zooming, and color modification based on relevant information. It takes the input from the User Interaction Layer and provides precise execution instructions for the underlying editing engine.

\textbf{Editing Engine Layer:} The Editing Engine Layer integrates LLM-based editing capabilities with the open-source SVGEdit framework, providing dual editing pathways: standardized function calling code generation through Agent Modifier, and direct manipulation through mouse/keyboard operations, with real-time information synchronization across modalities. The layer encapsulates core operations including element selection, transformation, and style modification, while supporting comprehensive SVG manipulation functions (creation, editing, rendering, export). Two-stage editing adaptation and cross-domain style compatibility ensure optimal efficiency across varying task complexities, with real-time preview enabling immediate visual feedback.

\textbf{Data Persistence Layer:} This layer maintains SVG templates across domains such as smart homes and sports and health, together with preset UI component libraries. It also records operation logs and uses interaction history to improve instruction parsing and personalize the editing experience.

\subsubsection{System Interface}
\begin{figure}[h]
  \centering
  \includegraphics[width=\linewidth]{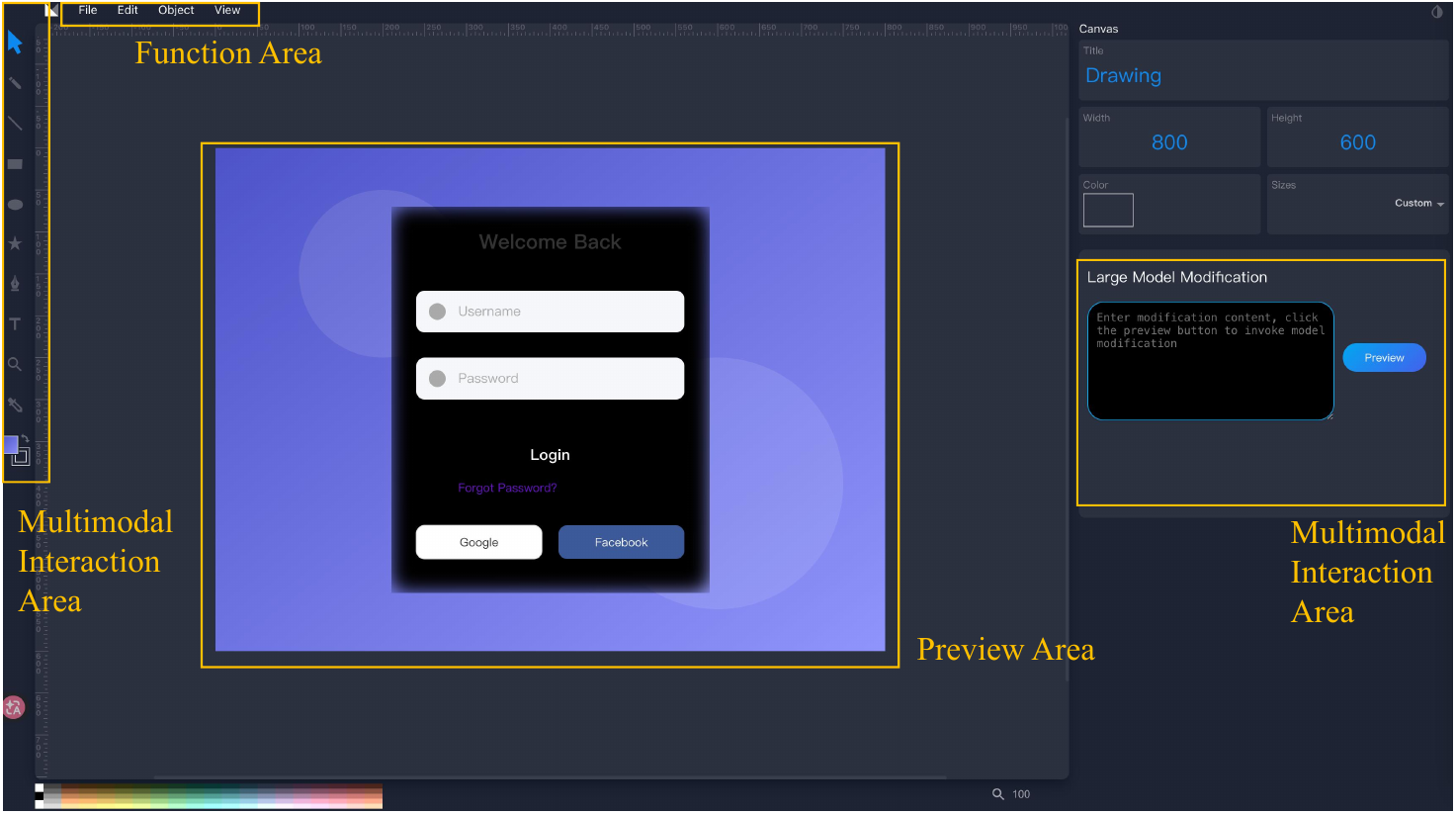}
  \caption{Interface of the multimodal SVG editing system.}
  \Description{Interface of the multimodal SVG editing system.}
  \label{fig:3}
\end{figure}

Based on the basic layout of SVGEdit, the main interface of this system adds a natural language input box and LLM function buttons, with real-time preview of editing effects in the SVG editing area. Users can enter natural language descriptions in the input box and click the confirm button to trigger functions such as intelligent generation and editing. The multimodal interaction entrance and cross-domain adaptation functions are enhanced. The interface is divided into three modules: the multimodal interaction areas on the left and right sides, the middle SVG editing and preview area, and the top preview and result area. An example diagram of the main interface of the system is shown in Figure~\ref{fig:3}.

The multimodal interaction areas on the left and right sides contain a natural language input bar and a mouse/keyboard operation toolbar. They support real-time display of element attributes, such as coordinates, color, and size, and update the attribute panel immediately after modification. The middle SVG editing area retains the original drawing tools of SVGEdit and adds a "multimodal operation prompt" function; selected elements are highlighted in purple. The top preview and result area includes an SVG source-code preview, download, and SVG file upload.

\section{Dataset Construction}
\label{sec:dataset}

To meet the research requirements for SVG element visual grounding and intelligent editing, existing public datasets cannot directly support the core tasks. Therefore, it is necessary to construct a structured and editable UI dataset in SVG format. As a vector format, SVG can accurately describe element attributes and structures through XML tags, serving as a key bridge between natural language interaction and large language model reasoning.

\begin{figure}[h]
  \centering
  \includegraphics[width=\linewidth]{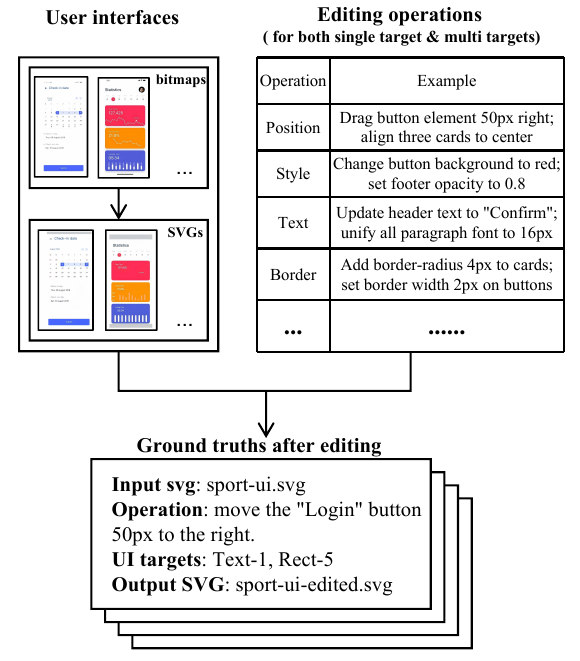}
  \caption{A diagram of dataset construction pipeline. 
  Bitmap-formatted UI screenshots are converted to SVG format. 
  Editing operations data is then constructed based on the operation taxonomy. 
  LLM processing generates ground truth annotations capturing operation descriptions, affected target elements, and edited SVG outputs. 
  The resulting dataset enables subsequent model fine-tuning and comprehensive evaluation.}
  \Description{Dataset construction pipeline showing UI-to-SVG conversion, editing operation taxonomy, LLM-based ground truth annotation generation, and downstream model fine-tuning and evaluation.}
  \label{fig:4}
\end{figure}

Our dataset construction follows a systematic pipeline (Figure~\ref{fig:4}). 
First, we convert bitmap-formatted UI screenshots to SVG representation, extracting and mapping UI element attributes to standardized vector formats. 
Second, we define a comprehensive taxonomy of 11 core UI editing operations covering common design practices. 
Third, we leverage LLM-based processing to automatically generate and filter instruction-target-output ground truth annotations at scale. 
This three-stage workflow progressively transforms raw bitmap UI data into a structured, semantically-rich dataset of 14,476 instruction-UI pairs that enables fine-tuning of visual grounding and editing models. 
The resulting dataset directly supports subsequent model training, evaluation, and system verification with minimal manual effort.

\subsection{Convert to SVGs}

Although the Rico dataset contains more than 66k mobile UI screens with semantic annotations and hierarchical structures, it is essentially based on bitmaps and annotations and lacks element-level editability. The mrtoy dataset~\cite{mrtoy2023mobileui} provides fine-grained annotations for UI element detection, such as bounding boxes, colors, and categories of texts and rectangles, but its native bitmap format cannot support precise element editing.

Compared with the Rico dataset, this work makes heavier use of the mrtoy dataset. Designed for mobile UI element detection tasks, this dataset covers target objects including text, images, and component groups, and provides UI images together with object detection bounding boxes along with their corresponding class labels and location information.

Bitmaps store image information as pixel matrices, so modifying a single UI element requires complicated pixel-level operations and cannot establish hierarchical semantic relationships between elements. To tackle these limitations, we extract core attributes of UI elements and map them to standard SVG vector formats under predefined rules, converting non-editable bitmap UI data into structured and semantically meaningful editable vector graphics.

In the dataset construction pipeline, we sequentially carry out UI element attribute extraction and SVG conversion, reconstruction of interface hierarchies and assignment of unique identifiers, validation of SVG specification compliance, and mobile-adaptive data filtering. The overall process maintains fidelity to the original data while satisfying the requirements of multimodal interaction and fine-grained editing. The final constructed SVG dataset fully preserves the original annotations, with clear UI hierarchies and unique element IDs, achieving an overall reconstruction accuracy of over 85\% that satisfies the needs of medium-complexity UI scenarios.

\subsection{Define Editing Operations}

Building upon the SVG-format UI dataset constructed above, we summarize 11 core categories of UI editing operations from common mobile interface design practices. We clarify the semantic description and practical application scenarios of each operation type, establishing a unified instruction specification for building question-answer (QA) pairs in the fine-tuning dataset. These operations range from basic element-level adjustments to advanced layout and stylistic modifications, comprehensively covering mainstream UI design scenarios.

\begin{table}[h]
\centering
\caption{Diverse types of UI editing operations}
\label{tab:ui_operations}
\resizebox{\linewidth}{!}{
\begin{tabular}{ll}
\toprule
\textbf{Type} & \textbf{Brief description} \\
\midrule
Position     & Move, align, center, distribute elements \\
Scaling           & Resize, scale, unify element dimensions \\
Style            & Modify color, transparency, appearance \\
Text              & Edit content, font, style, text attributes \\
Border          & Set borders, adjust corner radius \\
Visibility        & Show, hide, toggle element visibility \\
Layer          & Adjust stack order, manage element layers \\
Element        & Insert, remove, modify element quantity \\
Rotation            & Rotate angle, crop element content \\
Layout       & Arrange layout, grid, adaptive distribution \\
Advanced        & Conditional style, relative positioning \\
\bottomrule
\end{tabular}
}
\end{table}

This taxonomy helps the model accurately grasp the semantic logic and valid execution scope of different editing behaviors. Building upon these 11 core UI editing operation categories, we further expand the taxonomical details to construct two comprehensive tables: one for single-target UI editing operations and another for multi-target UI editing operations. For each major operation category, we conduct detailed sub-categorization and collect representative examples, ultimately establishing a fine-grained inventory of UI editing types and corresponding instruction examples. Concurrently, we systematically map the applicable scopes of these UI editing operations, revealing that individual UI graphics in our test dataset can accommodate as many as 38 completely distinct editing instructions. This comprehensive UI editing instruction data serves as a critical foundation for subsequently generating high-quality and comprehensive QA pairs for SVG graphics.

\subsection{Generate Ground Truths}

Based on our constructed UI editing instruction table, we employ a strategy where GPT-4o traverses all UI elements sequentially, then traverses UI editing instructions, and generates authentic UI editing instructions using each UI element's ID as ground truth according to our provided diverse instruction templates. Compared to the inverse approach of traversing questions and having the LLM generate correct answers, the scheme of treating element IDs as unique identifiers possesses significant advantages. Specifically, the difficulty of generating questions with ID as ground truth is substantially lower than requiring the LLM to directly locate target element IDs from natural language questions. Even in cases where the model makes errors, the resulting problems exhibit semantic unreasonableness rather than completely erroneous ID identification, thereby yielding data quality that far exceeds schemes with direct ID misidentification errors. We then perform automatic filtering using DeepSeek-R1 to remove low-quality, ambiguous, or semantically inconsistent samples, followed by manual inspection and evaluation to ensure instruction validity and annotation accuracy. Finally, we obtain 14,476 high-quality QA pairs that cover all 11 core operation categories.

Each sample in this dataset consists of three core components: a standard SVG-format UI graphic, a natural language editing instruction that targets one or more SVG elements, and the ground-truth ID(s) of the corresponding UI elements to be manipulated. A concrete example is the Sport App UI interface where the editing instruction "move the entire sports planning area up by 15px" corresponds to the target element Group-1, ensuring tight correspondence between questions and element IDs. This structured design enables end-to-end training for models to understand textual intents, locate target elements, and perform precise UI editing operations.

Building upon this instruction recognition dataset, we further extend our work scope by carefully selecting well-designed UI interfaces. For a total of 213 editing instructions across these interfaces, we employed a hybrid approach combining LLM generation with manual verification to produce authentic edited SVG files. This edited SVG dataset directly supports evaluation tasks in multimodal SVG editing, including natural language instruction alignment verification, UI element editing accuracy assessment, and incremental editing effect evaluation.

\section{Evaluation}

This section elaborates on the evaluation protocol specifically designed to verify the effectiveness and superiority of the proposed models and methods. The core goal of this evaluation is to comprehensively assess whether the two-stage framework, which integrates visual grounding enhancement and structured function calling, can effectively address the limitations of traditional LLM-driven SVG editing.

\subsection{Evaluation for Agent Locator}
We evaluate Agent Locator on our UI Visual Grounding Dataset, which covers common UI styles and element layouts for SVG editing. The evaluation focuses on visual grounding, a prerequisite for precise editing. We use the vanilla Qwen2.5-7B model and its fine-tuned version as the primary test subjects because the model is widely adopted in research and industry. We also include several popular LLMs to provide representative comparative benchmarks.

We assess the visual grounding performance of the evaluated models. To ensure experimental rigor, the dataset is split into training and test sets during the training phase. The test set is carefully curated to cover diverse UI styles and element arrangements. In total, 500 test samples are selected following two key principles: first, to encompass a broad range of UI design styles and avoid evaluation bias; second, to include all 11 core categories of UI editing instructions to guarantee comprehensive validation. 

Each test sample consists of three essential components: the original SVG file, a clear editing instruction, and the ground-truth ID of the corresponding target UI element. In the testing phase, the original SVG file and the editing instruction are fed into the model, which is only required to output the ID of the recognized target element. The visual grounding accuracy is then computed by comparing the model-predicted ID with the pre-annotated ground-truth ID.
\begin{table}[H]
  \caption{Accuracy comparison of locating single-target and multi-target using different models}
  \label{tab:single_multi_target_acc}
  \centering
  \setlength{\tabcolsep}{4pt}
  \begin{tabular}{lcc}
    \toprule
    Model  & Accuracy (single) & Accuracy (multi) \\
    \midrule
    DeepSeek-R1 & 82.2\% & 68.3\% \\
    DeepSeek-V3 & 88.6\% & 65.8\% \\
    GPT-4o & 88.1\% & 60.0\% \\
    Qwen2.5-7B (Base) & 54.9\% & 62.8\% \\
    Qwen2.5-7B (Finetuned) & \textbf{94.2\% (+39.3\%}) & \textbf{93.4\% (+30.6\%})\\
    \bottomrule
  \end{tabular}
\end{table}

Experimental results demonstrate that the fine-tuned Qwen2.5-7B model achieves an overall accuracy of 92\% on a mixed evaluation set containing both single-target and multi-target instructions, considerably outperforming the unoptimized base model and even surpassing several commercial large language models. This finding confirms that dedicated visual grounding fine-tuning effectively strengthens the UI element recognition and visual grounding ability of low-parameter base models without prior task-specific training, laying a solid foundation for subsequent high-precision SVG editing.

\subsection{Evaluation for Agent Modifier}

We evaluated the performance of the function calling framework in Agent Modifier. The evaluation focuses on two core dimensions: editing accuracy and operational efficiency. We selected 5 mainstream models for comparative testing, including DeepSeek-V3.2, Claude-3-Haiku, Qwen3-32B, GPT-4o and Gemini 2.5 Flash. These models cover both commercial and open-source options, ensuring the generalizability of the experimental results.

In terms of editing accuracy, we compared the performance of the function calling method with that of direct editing. The results show that the function calling framework outperforms direct editing across all models. For instance, DeepSeek-V3.2 achieves an accuracy of 97.2\% under the function calling mode, whereas its accuracy is 89.7\% in the direct editing mode. This gap indicates that the structured function calling paradigm effectively reduces the randomness of model outputs and produces editing results that better align with user intentions.

\begin{table}[h]
  \caption{Modification accuracy comparison of direct prompting and function calling using different models}
  \label{tab:svg_acc}
  \centering
  \begin{tabular}{lcc}
    \toprule
    Model & Accuracy (direct) & Accuracy (FC) \\
    \midrule
    DeepSeek-V3.2 & 89.7\% & 97.2\% (+7.5\%) \\
    GPT-4o & 90.6\% & 95.5\% (+4.9\%) \\
    Gemini 2.5 Flash & \textbf{96.1}\% & \textbf{98.6}\% (+2.5\%) \\
    Claude-3-Haiku & 85.5\% & 87.8\% (+2.3\%) \\
    Qwen3-32B & 76.1\% & 87.3\% (+11.2\%) \\
    \bottomrule
  \end{tabular}
\end{table}

To evaluate operational efficiency, we analyze two critical metrics: token consumption and response time. As summarized in Table~\ref{tab:svg_token_time_transpose}, the function calling (FC) framework achieves substantial performance gains over direct prompting across both indicators. For the lightweight model Claude-3-Haiku, token consumption is reduced from 2010 to 178, representing a 91.1\% decrease, while response time drops from 18.8 seconds to 2.6 seconds, an 85.9\% reduction that supports real-time SVG editing. For Qwen3-32B, a representative open-source model, token consumption is cut from 2517 to 706 (72.0\% lower), and response time is shortened from 71.3 seconds to 21.3 seconds (70.1\% faster), significantly enhancing the practicality of the system in real-world deployment.

Function calling also makes LLM-based editing less opaque. Unlike direct generation, which can produce unpredictable outputs, its structured workflow yields more consistent results and reduces incorrect element selection and unintended modifications. This improves both reliability and practical usability.

\begin{table}[h]
  \centering
  \caption{Token consumption and response time of direct prompting and function calling}
  \label{tab:svg_token_time_transpose}
  \setlength{\tabcolsep}{4pt}
  \begin{tabular}{lcccc}
    \toprule
    \multirow{2}{*}{Model}
      & \multicolumn{2}{c}{Token consumption}
      & \multicolumn{2}{c}{Response time (sec)} \\
    \cmidrule(lr){2-3} \cmidrule(lr){4-5}
      & Direct & FC & Direct & FC \\
    \midrule
    Claude-3-Haiku
      & 2010 & 178 (-91.1\%)
      & 18.8 & 2.6 (-85.9\%) \\
    Qwen3-32B
      & 2517 & 706 (-72.0\%)
      & 71.3 & 21.3 (-70.1\%) \\
    \bottomrule
  \end{tabular}
\end{table}

 This evaluation also reveals several limitations of the proposed framework. The current pipeline relies on element IDs embedded in SVG files; in the absence of such IDs, additional preprocessing steps are required, which increases the overall system complexity. Furthermore, our experiments are conducted only on structured UI layouts, and the model performance in other practical scenarios—such as irregularly arranged elements—remains to be verified.
 
In summary, the two-stage framework proposed in this paper effectively integrates enhanced visual grounding with structured function calling, and significantly improves the accuracy and efficiency of LLM-driven SVG editing. By addressing the core drawbacks of conventional methods, it provides a reliable and practical solution for UI design and vector graphics editing. The framework not only ensures precise and stable editing operations, but also enhances the interpretability and controllability of the system, allowing users to directly participate in the LLM-based modification process. It therefore carries important practical value for real-world UI development and design workflows.

\section{Discussion}

This work addresses critical bottlenecks in multimodal-driven SVG editing systems, including modal fusion, low LLM editing accuracy, inefficient inference, and lack of controllability over model generation. We propose a multimodal-driven SVG editing method and implement a complete SVG editing system. For LLM-based editing, we design a two-stage pipeline that integrates specialized visual grounding fine-tuning with a structured function calling paradigm, validating its superiority through extensive quantitative and qualitative experiments.

Dedicated fine-tuning for visual grounding significantly enhances the model's UI element visual grounding capability. By optimizing the Qwen2.5-7B model on our custom UI-oriented dataset, the overall accuracy on a mixed evaluation set containing both single-target and multi-target instructions is improved from 62\% to 92\%, surpassing all tested commercial LLMs. As the fundamental prerequisite for accurate SVG manipulation, this breakthrough eliminates the core limitation of weak visual perception in general LLMs, ensuring that all subsequent editing operations are performed on the correct target elements.

Built upon robust visual grounding, our structured function calling framework optimizes the entire SVG editing workflow. Comprehensive evaluations on editing accuracy and operational controllability demonstrate that our approach consistently outperforms the native direct generation method across mainstream open-source and commercial LLMs. For accuracy, the framework elevates the performance of most models, such as boosting Qwen3-32B from 76.1\% to 87.3\% (improvement of +11.2\%). Regarding controllability, our method enables users to participate in the model's editing process more effectively, ensuring stable, consistent, and user-intent-aligned editing results through a tool-guided pipeline, rather than producing unpredictable and error-prone outputs.

Despite introducing more complex system design, our method significantly reduces token consumption and response time. On lightweight models, response time is reduced by 85.9\% and token consumption by 91.1\%; on large open-source models, we achieve a 70\% reduction in latency and corresponding reductions in token consumption. These substantial efficiency improvements significantly enhance model editing efficiency, making real-time and cost-effective UI editing feasible.

Although this work achieves notable results, several limitations warrant further investigation. First, the system is designed and trained primarily on SVG-based UIs, and its performance on other domains or more complex SVG graphics remains to be validated. Second, large state-of-the-art LLMs already provide strong editing capabilities. For example, Gemini 2.5 Flash achieves 96.1\% accuracy, so our method improves its accuracy by only 2.5\%. Nevertheless, our method remains valuable because it reduces reasoning time and token consumption while improving response speed.

Looking forward, the recent paradigm shift from function calling to skill-based abstractions 
marks a significant upgrade in LLM agent design. Although this work currently employs 
structured function calling for SVG editing, it is noteworthy that skill-based approaches, 
while encapsulating tool interactions and autonomous decision-making as skills, still 
fundamentally depend on structured function calling at their core. Therefore, the function 
calling framework proposed in this work provides a foundational basis for future skill-based 
abstractions, ensuring that our approach remains applicable as agent paradigms continue to 
evolve.

\section{Conclusion}

This work addresses critical bottlenecks in multimodal SVG editing systems—modal fusion, 
low editing accuracy, inefficient inference, and insufficient controllability—by proposing 
a comprehensive solution. We introduce a two-stage strategy—visual grounding first, then 
modification—in which both stages support dual interaction modalities. Through this design, 
we have successfully implemented a high-precision, efficient, and controllable SVG editing 
system. Our work makes the following contributions: We construct a UI domain-specific SVG dataset 
that enables systematic training and evaluation of LLMs. We propose a two-stage strategy 
that effectively addresses the long-standing challenge of UI element visual grounding in 
LLMs, significantly enhancing editing accuracy and user controllability while substantially 
reducing computational costs. We design MM-SVGEdit, a multimodal SVG editing system that 
integrates multimodal-driven visual grounding and structured editing into a coherent 
pipeline. These contributions provide a practical solution for structural understanding 
and manipulation of user interfaces, advancing the capability of LLMs in handling complex 
UI editing tasks.

\bibliographystyle{ACM-Reference-Format}
\bibliography{references}

\end{document}